\documentclass[referee]{raa}            
\usepackage{graphicx,times}             
\input{epsf.sty}                        
\input{psfig.sty}                       

\begin{document}

   \title{The causality between the rapid rotation of a sunspot and an X3.4 flare
}

   \volnopage{Vol.0 (200x) No.0, 000--000}      
   \setcounter{page}{1}          

   \author{Xiao.-Li. Yan
      \inst{1,2}
   \and Zhong-Quan. Qu
      \inst{1}
   \and Cheng-Lin. Xu  \inst{3} \and Zhi-ke. Xue  \inst{1,2} \and De-Fang, Kong  \inst{4}
   }

   \institute{National Astronomical Observatories/Yunnan Observatory, Chinese Academy of Sciences,
             Kunming 650011, China; {\it yanxl@ynao.ac.cn}\\
        \and
             Graduate School of Chinese Academy of Sciences, Beijing, China\\
        \and
             Yunnan Normal University, Kunming Yunnan, China\\
             \and
             Jiaxing University, Jiaxing Zhejiang, China\\
   }

   \date{Received~~2009 month day; accepted~~2009~~month day}

\abstract{ Using multi-wavelength data of Hinode, a rapid rotation
of a sunspot in active region NOAA 10930 is studied in detail. We
found the extra-ordinary counterclockwise rotation of the sunspot
with positive polarity before an X3.4 flare. From a series of vector
magnetograms, it is found that the magnetic force lines highly
sheared along the neutral line accompanying the sunspot rotation.
Furthermore, it is also found that the sheared loops and an inverse
S-shaped magnetic loop in the corona formed gradually after the
sunspot rotation. The X3.4 flare can be reasonably regarded as a
result of this movement. The detailed analysis provides an evidence
that sunspot rotation leads to magnetic field lines twisting in the
photosphere, the twist is then transported into the corona, and
finally flares are triggered.  }

   \authorrunning{Xiao.-Li. Yan, Zhong-Quan. Qu, Cheng-Lin. Xu, Zhi-ke. Xue, De-Fang, Kong }            
   \titlerunning{The causality between the rapid rotation of a sunspot and an X3.4 flare }  

   \maketitle

%
%
\section{Introduction}           
\label{sect:intro}

It is now widely accepted that flares derive their power from the
free energy stored in stressed or non-potential magnetic fields in
the active regions (Zirin et al. 1973; Hagyard et al. 1984). But the
process of rapid transformation of the magnetic energy into kinetic
energy of particles, radiation, plasma flows and heat has not been
very clear in detail. How the magnetic energy is stored and released
still needs more observational evidences. Thus it remains a very
important issue in solar physics.

There are several characteristics of active regions which are in
favor of causing flares. If the active regions have strong magnetic
gradients (Wang, et al. 1994, 2006), highly sheared transverse
magnetic fields (Rausaria et al. 1993; Chen et al. 1994; Wang et al.
1996), emerging fluxes (Schmieder et al. 1997; Chen \& Shibata
2000), and flux cancellation (Wang \& Shi 1993; Zhang \& Wang 2001;
Zhang et al. 2001), flares are more often triggered. Besides the
observations mentioned above, sunspot rotation (Evershed 1910,
Nightingale et al. 2000, 2002; Brown et al. 2003, Yan et al. 2008)
may be involving with energy buildup and the energy is released
later via flares (Stenflo, 1969; R$\acute{e}$gnier et al. 2006; Yan
et al. 2007, 2008). Rotation of a sunspot in the photosphere may
cause an injection of twist into the corona (Tian et al. 2006),
which was proven by the TRACE EUV images, also by the S-shaped or
inverse S-shaped structures in the soft X-ray images of Yohkoh/SXT
(Canfield et al. 1999, Pevtsov 2002).

In this paper, we describe the rapid rotation of the sunspot and the
relationship between the sunspot rotation and the X3.4 flare. An
emphasis on the possible causality of the flare eruption is made by
tracing the evolutions from the photosphere to the corona of this
active region. This may add one significant and reliable evidence to
an argument that the rotation motion trigger the flare.

\section{Observational data}
\label{sect:Obs}

The data used in this paper contain those detailed as follows: 1.
Continuum intensity images and vector magnetograms from
Spectropolarimeter (SP) of Solar Optical Telescope (SOT, Tsuneta et
al. 2008), and X-ray images from the X-Ray Telescope (XRT, Golub et
al. 2007) aboard on Hinode (Kosugi et al. 2007); 2. Soft X-ray flux
of GOES12.

In this paper, we used the Fast Map of SP data. The reduction of
spectropolarimetric data from Hinode is carried out by the radiative
transfer equation derived by Unno (1956) and improved by Landolfi et
al. (1982), and a Milne-Eddington atmosphere is assumed. The 180
degree ambiguity is resolved by comparing vector magnetograms with
the potential fields (Metcalf et al. 2006).

\begin{figure*}
\centering
 \includegraphics[width=2.5in]{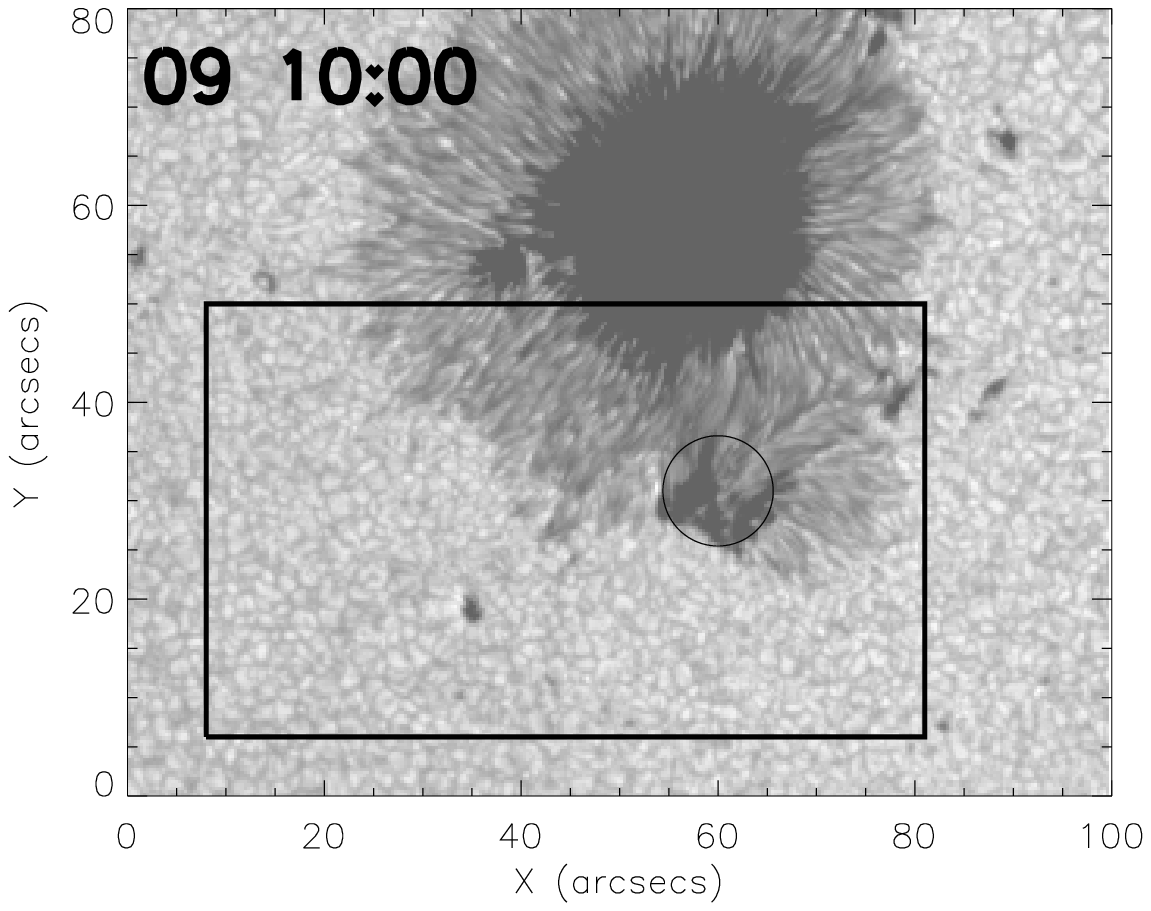}
 \includegraphics[width=2.5in]{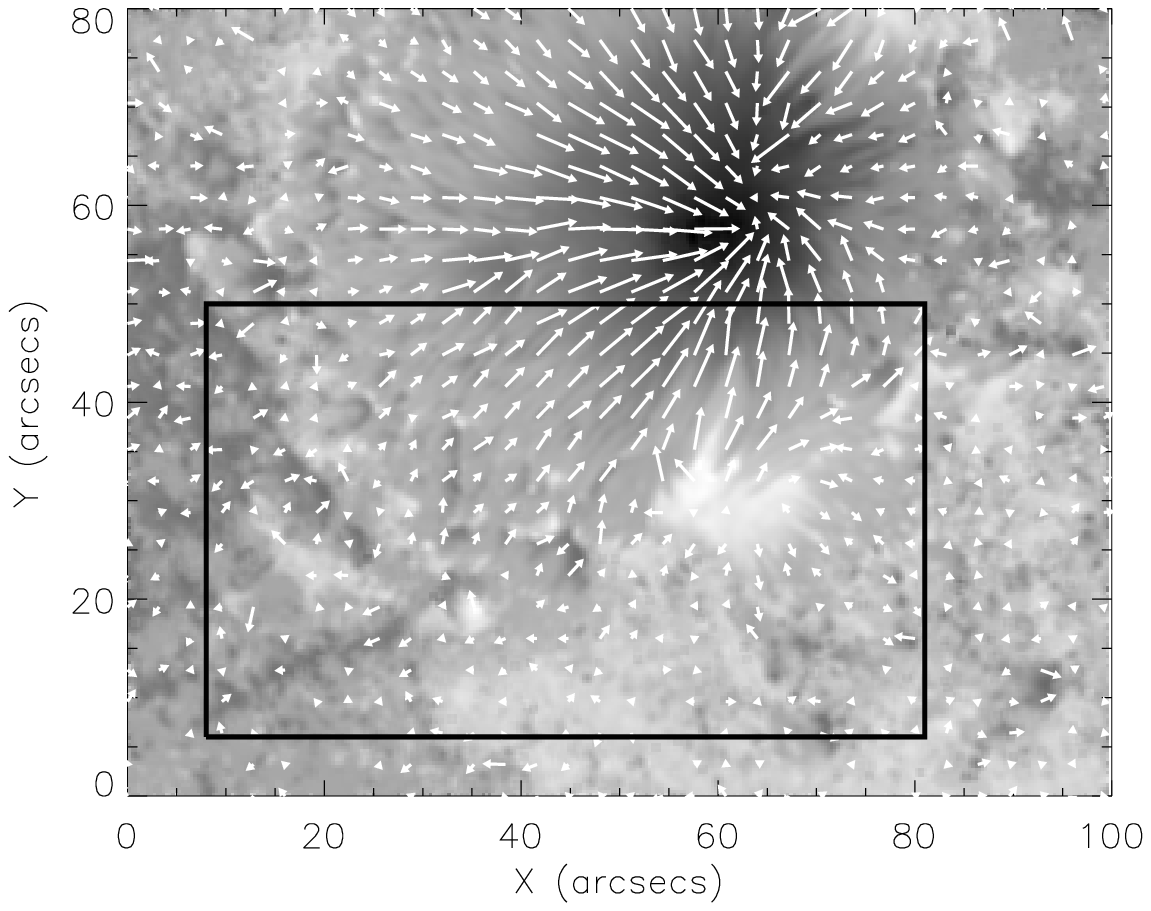}\\
\caption{The continuum intensity image (left panel) and vector
magnetogram (right panel) of high resolution processed from
Spectropolarimeter of SOT on 2006 December 9 respectively. Black
(white) patches in the right panel indicate the negative (positive)
polarity. The circle is placed at the rotating sunspot and the box
outlines the field of view of Fig. 3.} \label{}
\end{figure*}

\begin{figure}
\centering
\includegraphics[width=7cm]{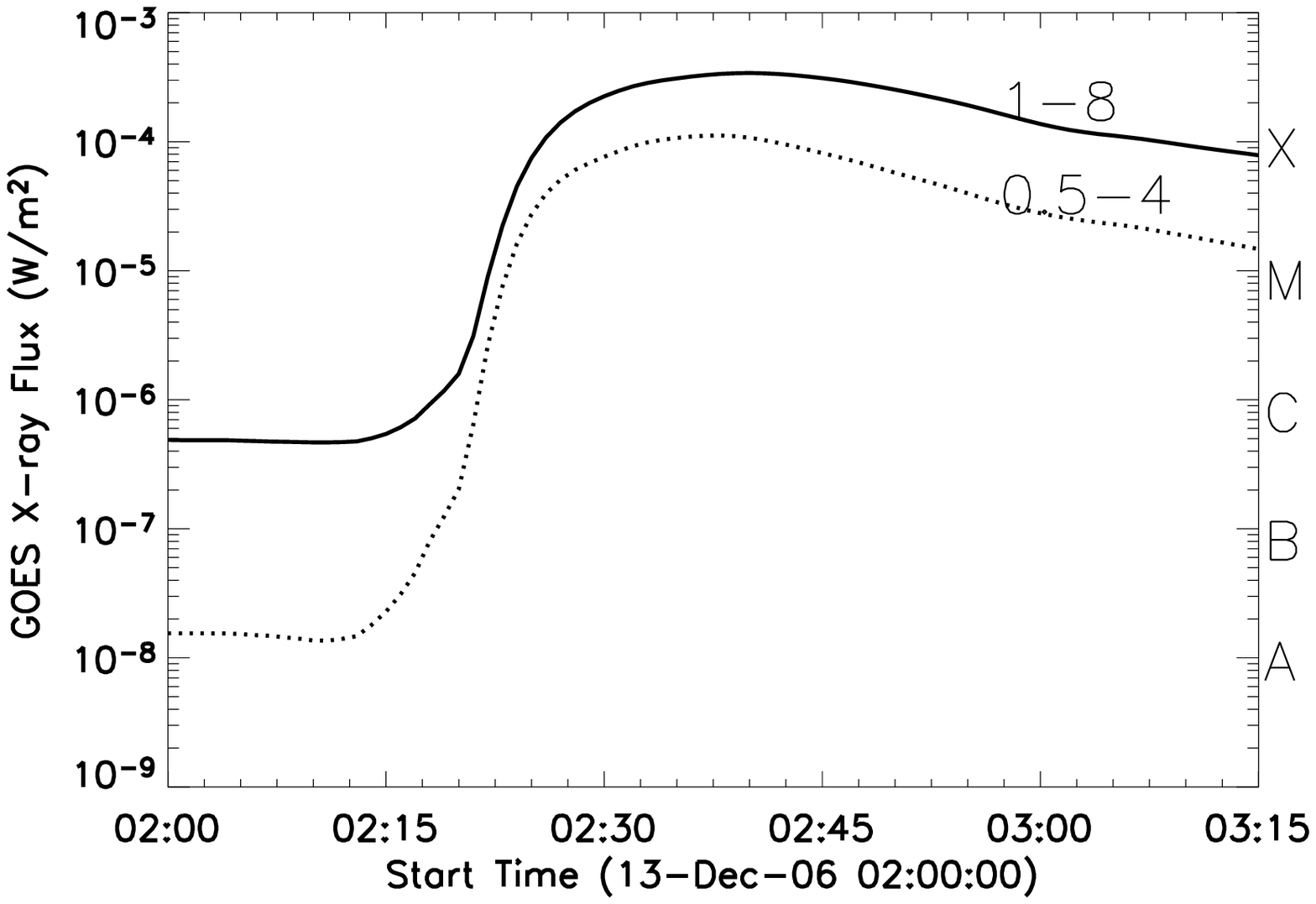}
\caption{Evolution of soft X-ray emission (Solid line: 1-8 \AA.
Dashed line: 0.5-4 \AA) for the flare on 2006 December 13 from GOES
12.} \label{}
\end{figure}

\begin{figure*}
\centering
  \includegraphics[width=1.6in]{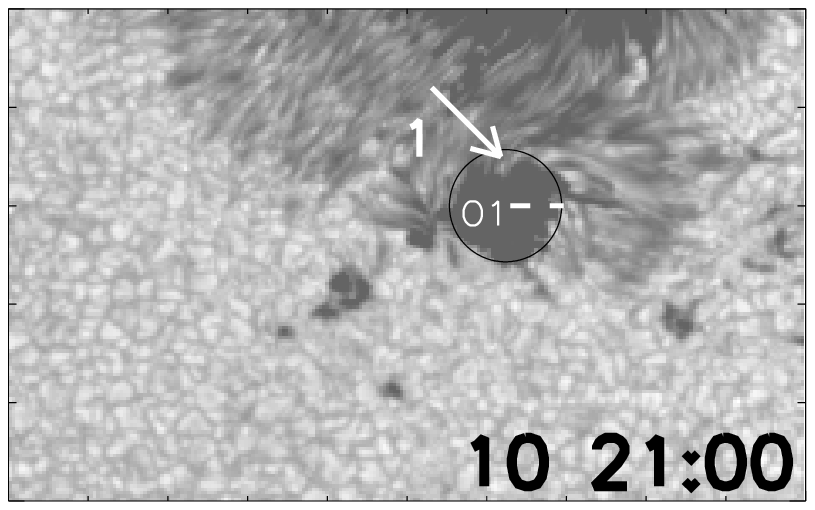}%
  \includegraphics[width=1.6in]{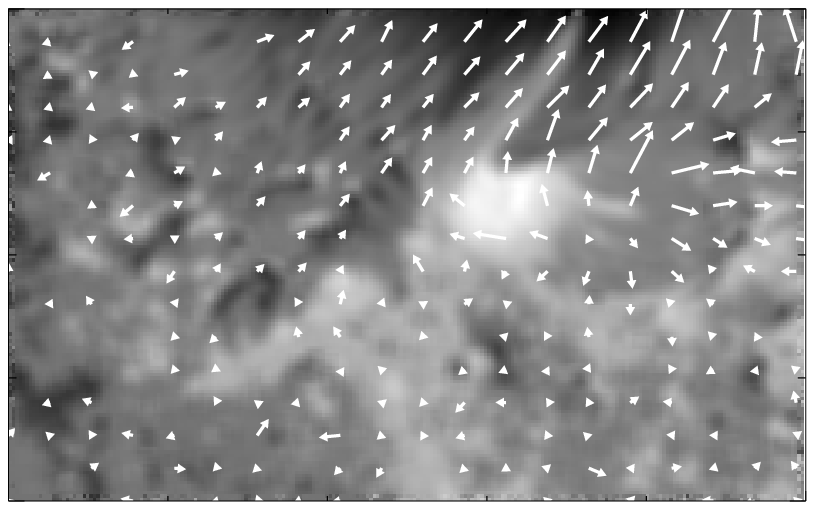}\\
  \includegraphics[width=1.6in]{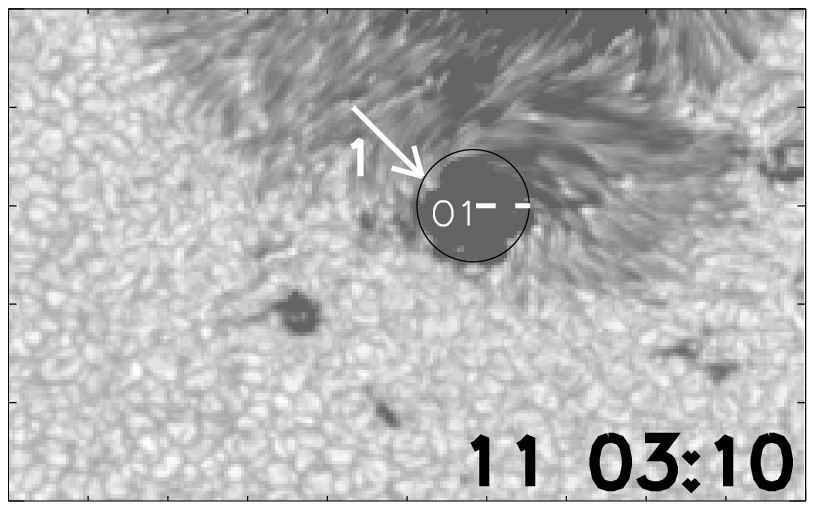}%
  \includegraphics[width=1.6in]{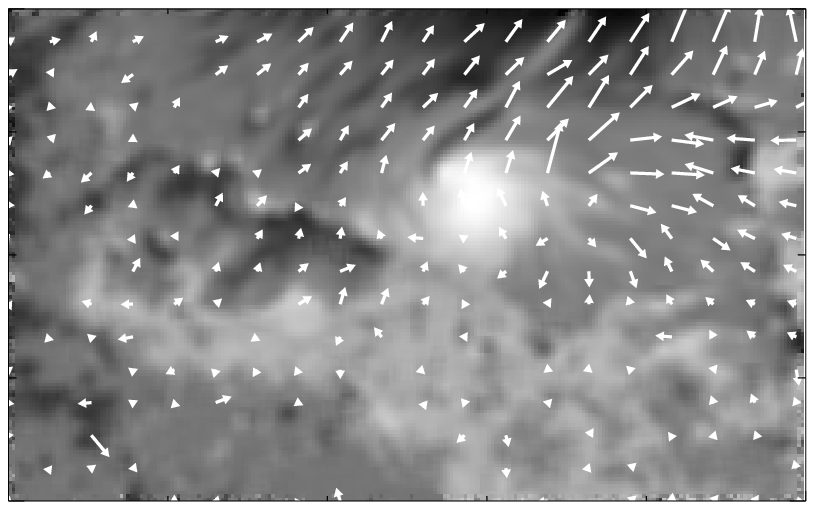}\\
  \includegraphics[width=1.6in]{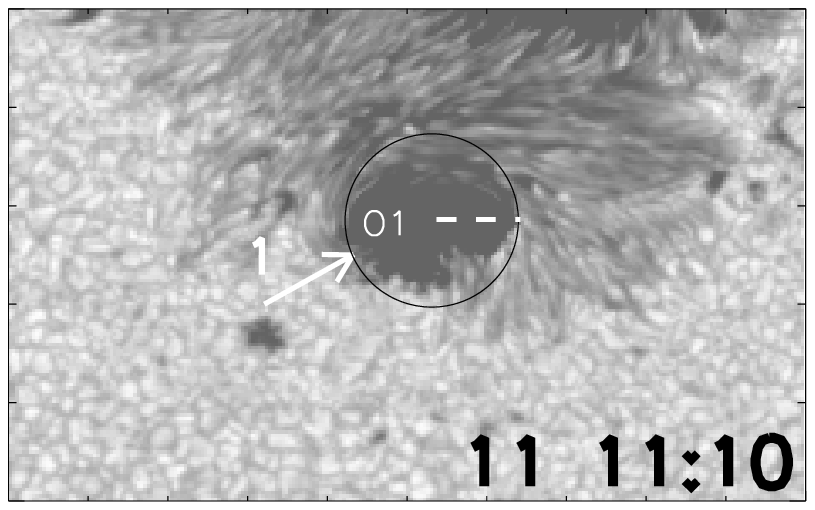}%
  \includegraphics[width=1.6in]{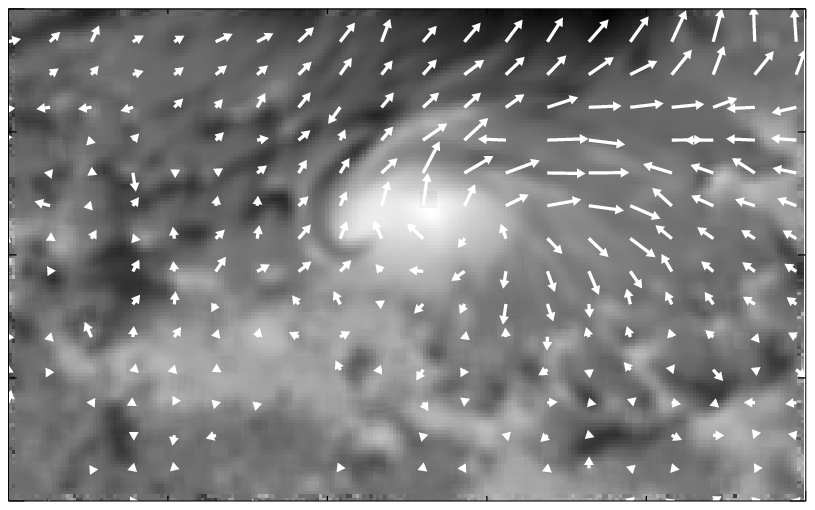}\\
  \includegraphics[width=1.6in]{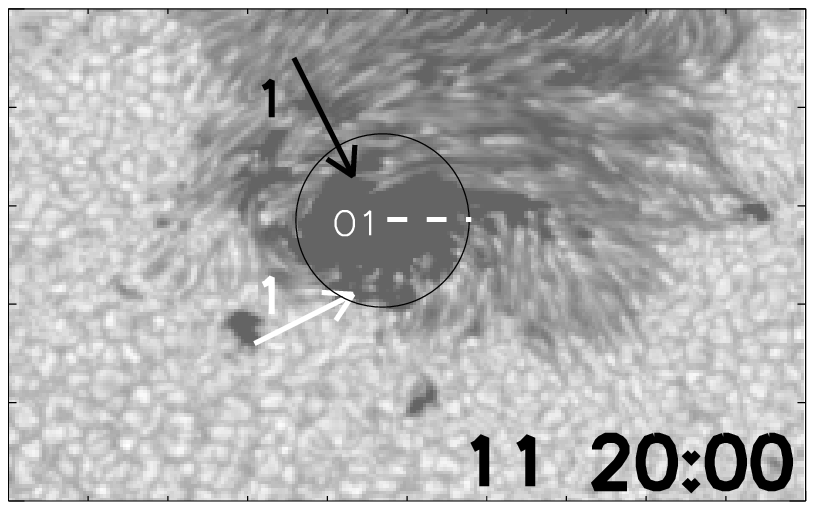}%
  \includegraphics[width=1.6in]{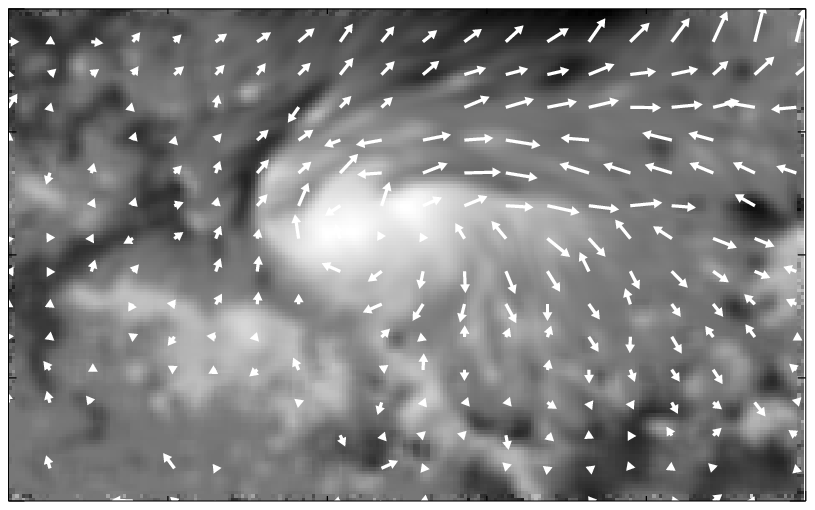}\\
  \includegraphics[width=1.6in]{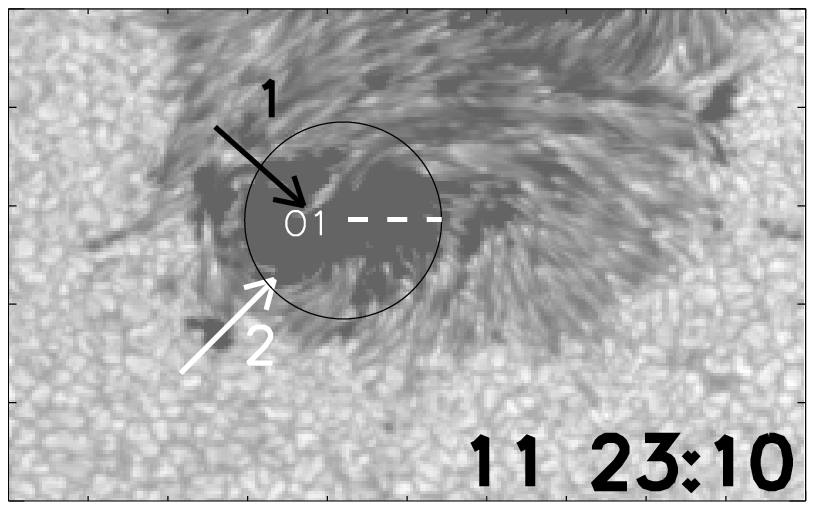}%
  \includegraphics[width=1.6in]{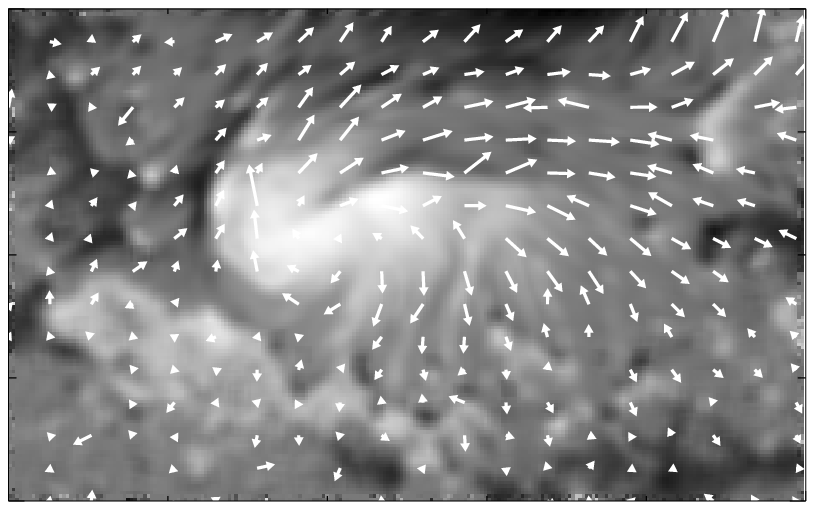}\\
  \includegraphics[width=1.6in]{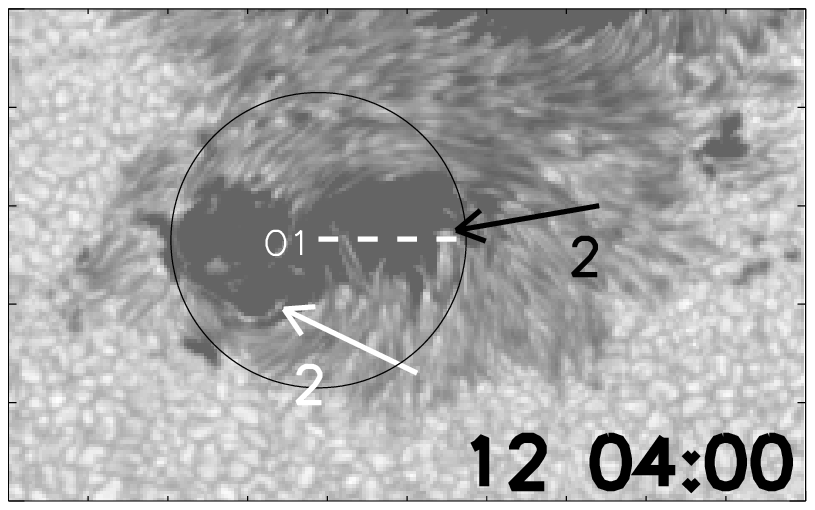}%
  \includegraphics[width=1.6in]{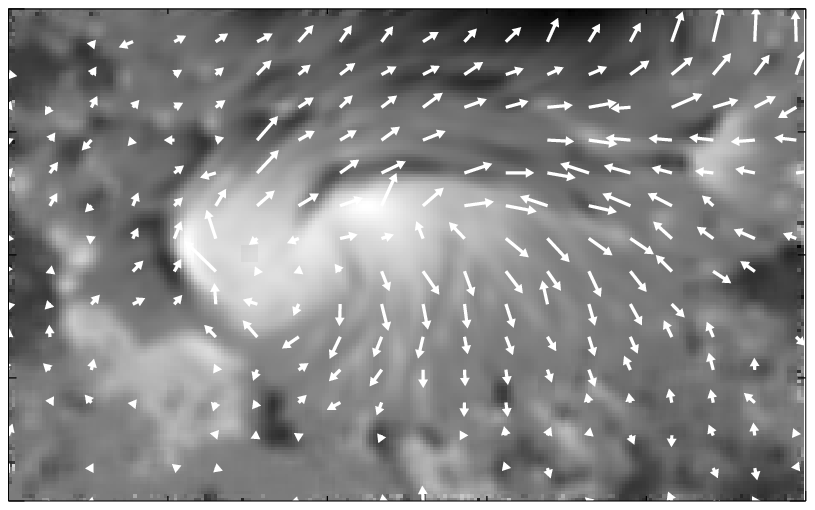}\\
  \includegraphics[width=1.6in]{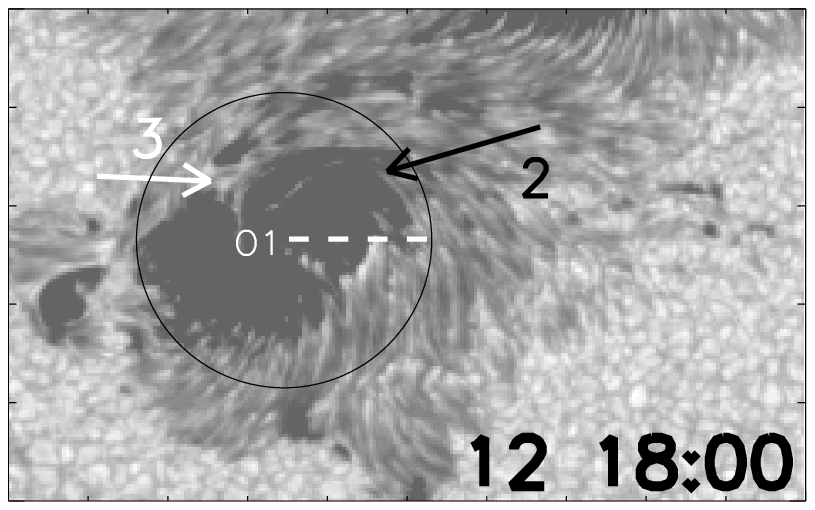}%
  \includegraphics[width=1.6in]{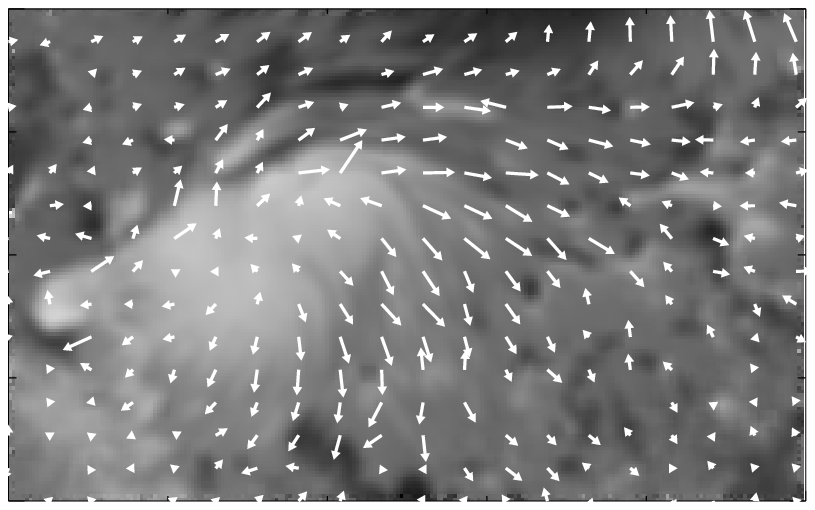}\\
  \includegraphics[width=1.6in]{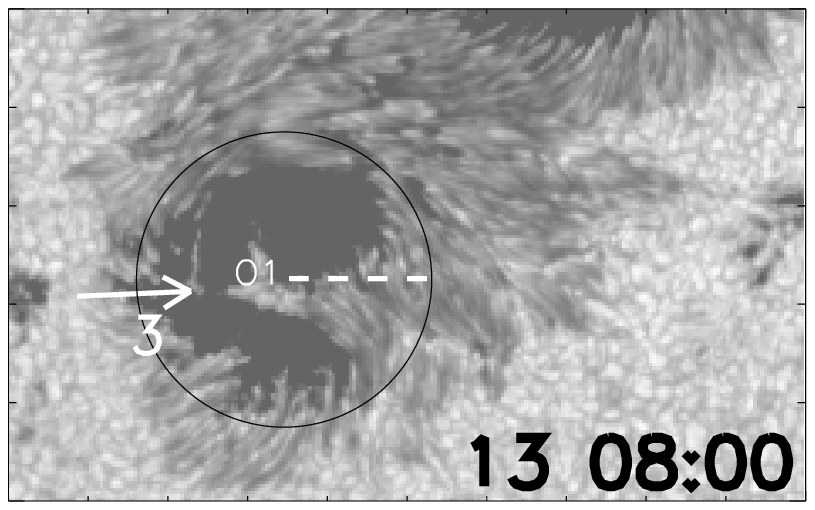}%
  \includegraphics[width=1.6in]{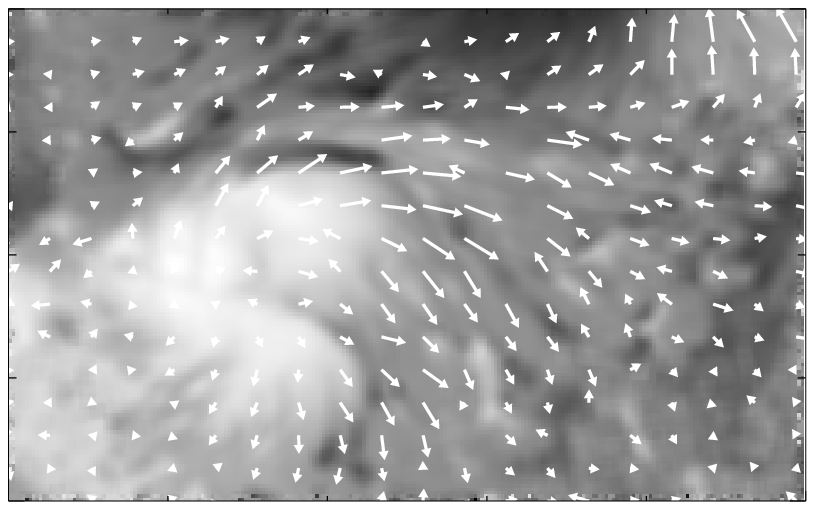}\\
\caption{Rapid rotation of sunspots seen in the sequence of the
continuum intensity images and vector magnetograms from
Spectropolarimeter of SOT. The circle is including the umbra of the
rotating sunspot. The arrows in the left panels are specified in
detail in the text. The field of view is 75$^\prime$$^\prime$
$\times$ 45$^\prime$$^\prime$.} \label{}
\end{figure*}

\begin{figure*}
\centering
 \includegraphics[width=5in]{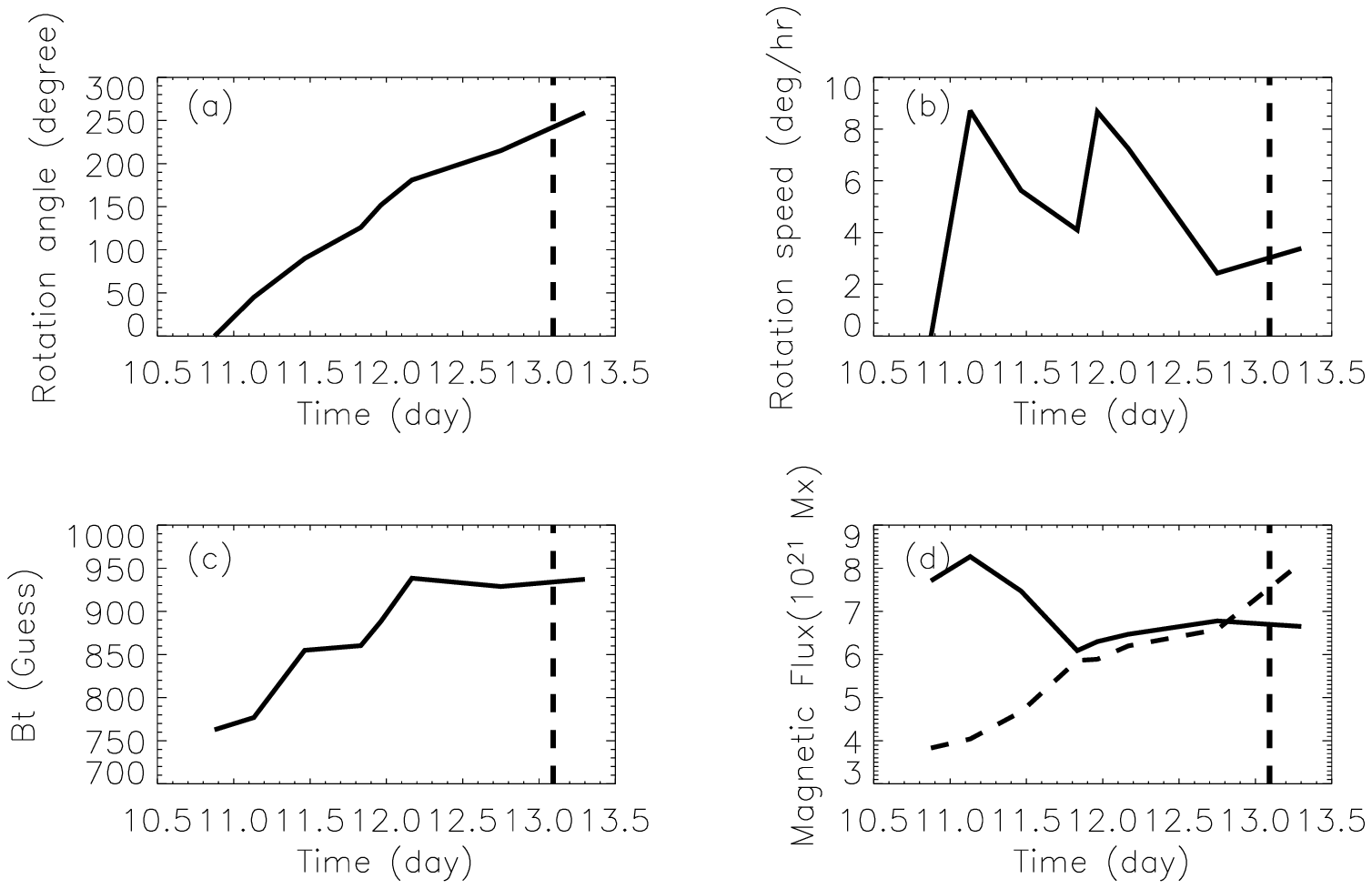}
\caption{Plots showing the evolution of the rotation angles (a),
rotation speed (b), the average strength of transverse magnetic
fields (c), and the positive (solid line)/negative (dashed line)
magnetic fluxes (d). The vertical dashed lines denote the beginning
time of the X3.4 flare.} \label{}
\end{figure*}

\begin{figure*}
\centering
 \includegraphics[width=2in]{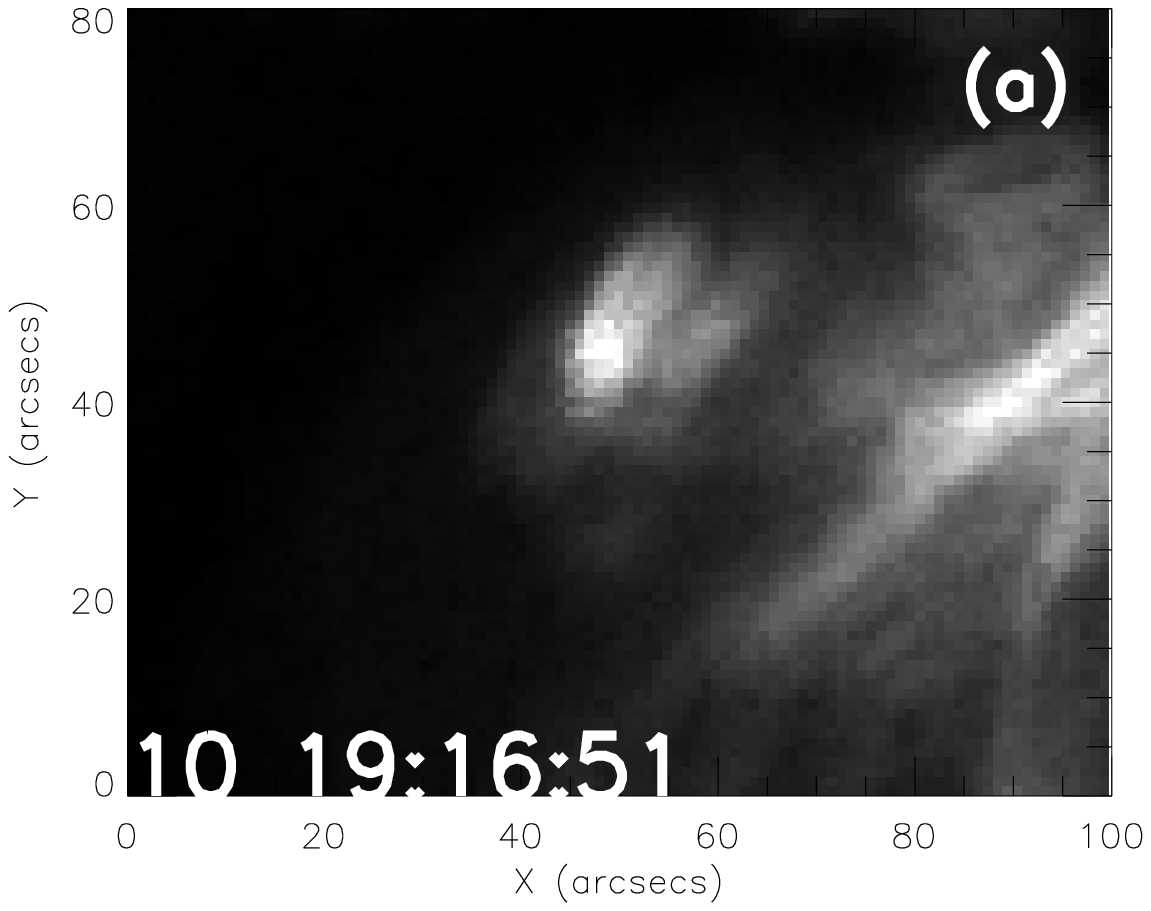}%
  \includegraphics[width=2in]{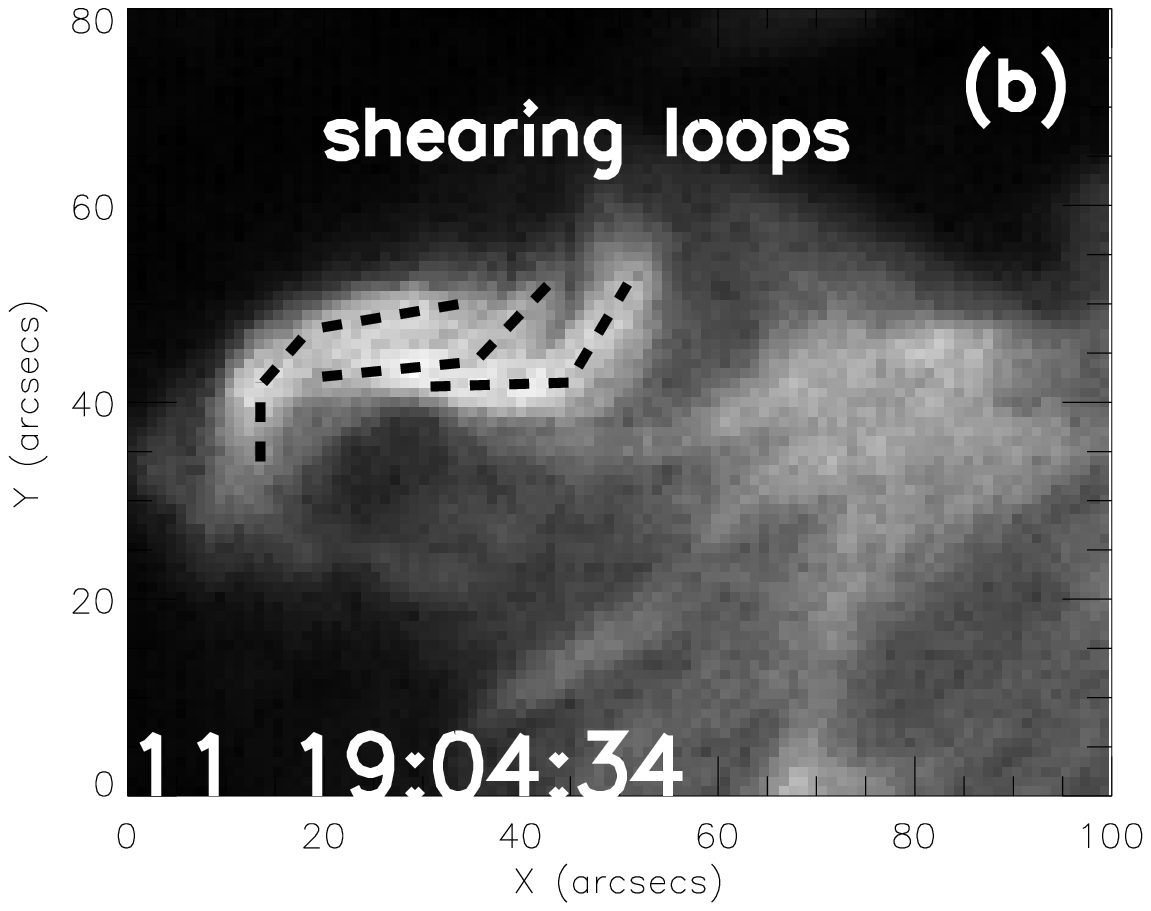}\\
    \includegraphics[width=2in]{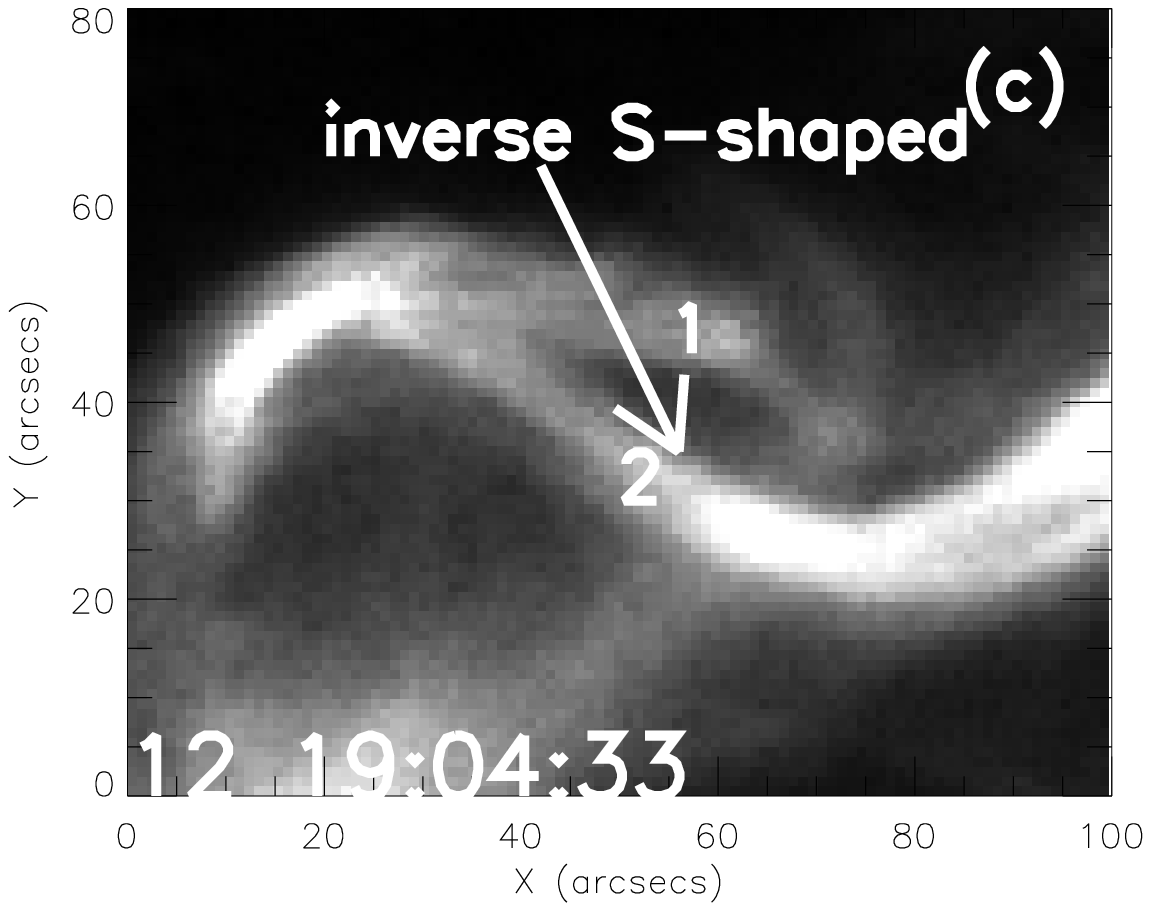}%
  \includegraphics[width=2in]{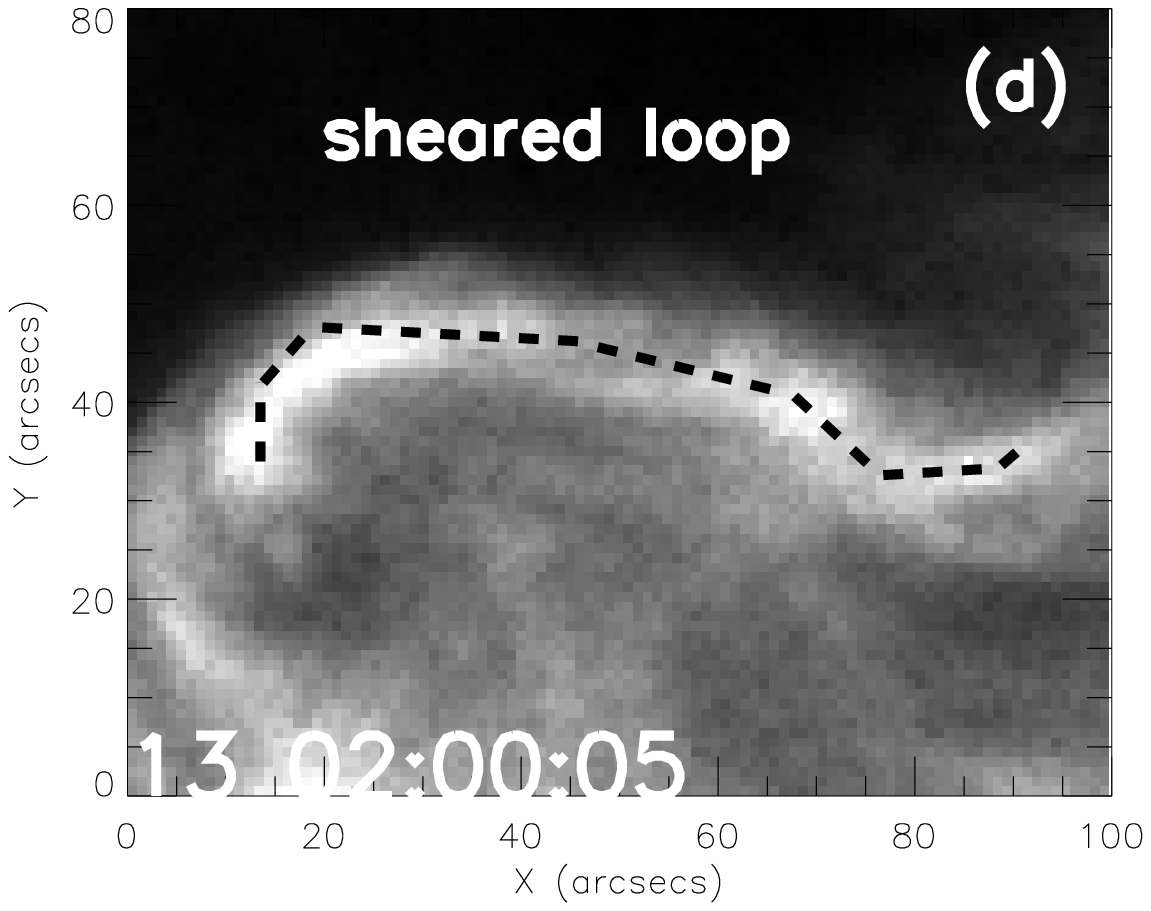}\\
    \includegraphics[width=2in]{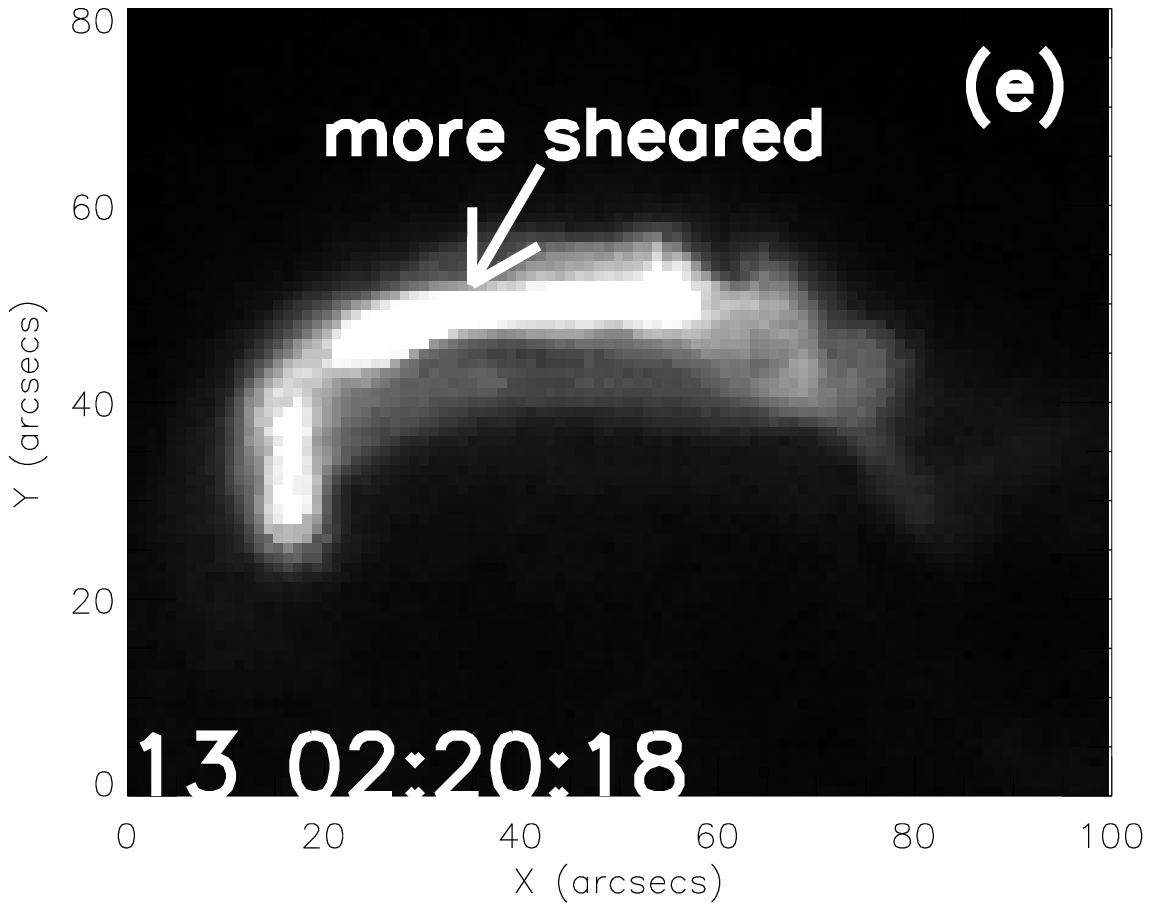}%
  \includegraphics[width=2in]{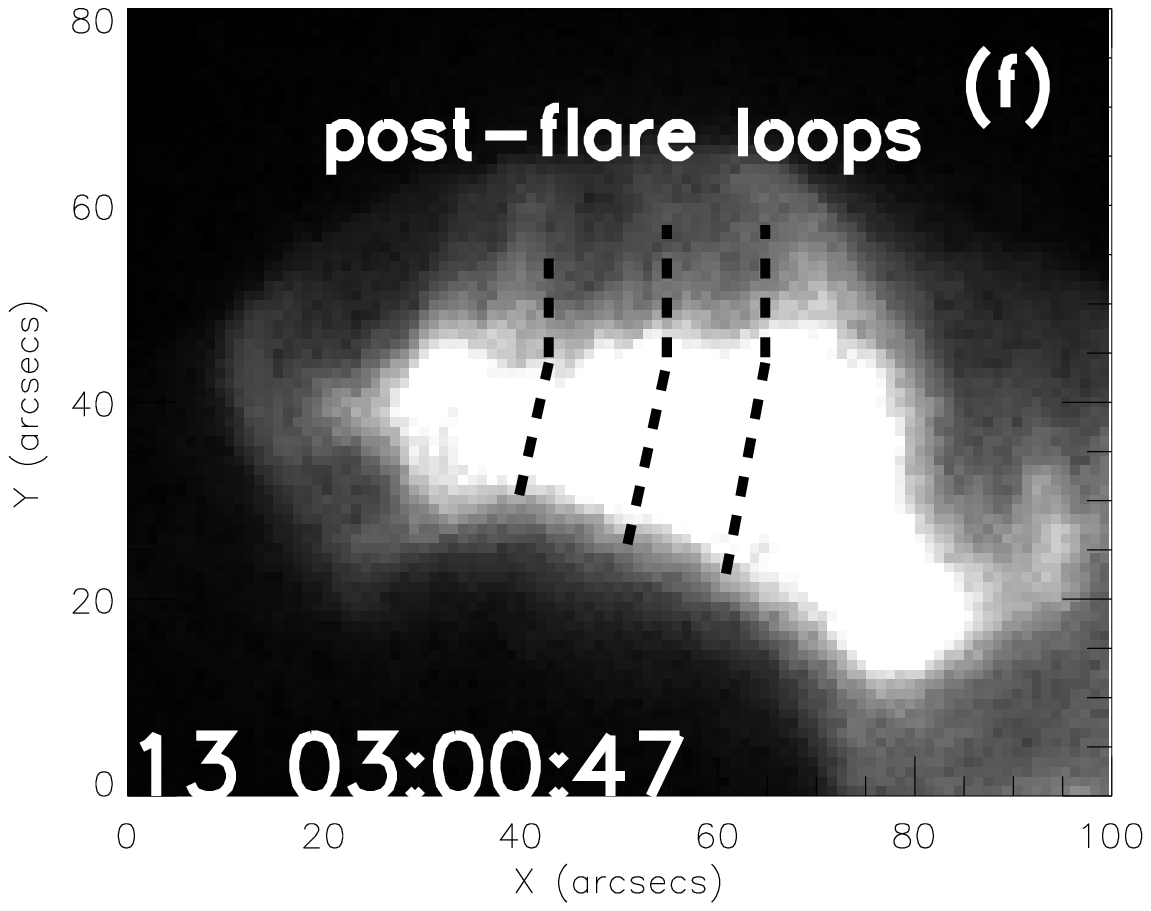}\\
\caption{A series of X-ray images observed with Be-thin filter by
XRT of Hinode from 2006 December 10 to December 13. The dashed lines
and the arrows are described in the text.} \label{}
\end{figure*}

\section{Process of an X3.4 flare}
\label{sect:flare}

NOAA 10930 is a bipole active region, which was composed of a big
sunspot with negative polarity and another small sunspot with
positive polarity (see Fig. 1). There were many B class flares, C
class flares, and two X class flares occurring in this active region
recorded from December 9 to 14. Here, we only analyze the prominent
X3.4 flare in detail for the purpose.

According to the soft X-ray emission from GOES 12 (see Fig. 2),
there was an X3.4 flare in this active region. It occurred at
02:14UT and reached its peak at 02:40UT.

By observing the evolution of this active region, we noted that the
small sunspot presented an extra-ordinary counterclockwise rotation.
Figure 3 shows high resolution continuum intensity images (left
panels) and vector magnetograms (right panels) with transverse
components overlying longitudinal ones gotten from
Spectropolarimeter of SOT. From the evolution of the continuum
intensity images, one can see that the small sunspot with positive
polarity not only moved from the southwest of the big sunspot to the
southeast but also rotated rapidly around its umbral center.
Referring to the light fibrils connecting the umbra of the small
sunspot in the continuum images, one can trace its rotation process.
The center of the circle which contains all the umbral features of
the small sunspot is labeled as $``O1"$ to illustrate the reference
frame. The light fibrils also rotated around the center of the
sunspot umbrae, one can evaluate their motions and obtain rotation
angle and speed easily. Arrows labeled by numbers indicated
different penumbral fibrils which are used for tracing the change of
the rotation. In this figure, the dashed lines in the small sunspot
indicate the radius of the circle. The rotation angle increased
during the evolution of four days and reached 259 degrees (see Fig.
4a). Two peaks appeared in the curve recording the rotation speeds
on Dec. 11, 12 (see Fig. 4b). Because the time resolution is not so
high, we only get relatively rough average of the rotation speed.
Following the sunspot rotation, magnetic force lines between the two
sunspots around the magnetic inversion line became sheared
correspondingly (see the right panels of Fig. 3). The transverse
magnetic fields took on spiral pattern around the center of the
umbrae of the rotating sunspot. The average strength of transverse
magnetic fields in the box in Fig. 1 increased rapidly before
December 12 (see Fig. 4c). Simultaneously, the expansion of the
magnetic field-covered area (the box labeled region in Fig. 1)
related to the rotating sunspot can be seen. The negative magnetic
fluxes increased from December 11 to 13. However, the positive
magnetic fluxes first increased and then decreased rapidly (see Fig.
4d). After we checked the magnetograms, we found that there are two
reasons accounting for this phenomenon. On one hand, there are many
samll positive magnetic fluxes around the main positive sunspot
diffused out of the region following the main positive sunspot
emergence. On the other hand, some of the longitudinal magnetic
fields may be transfered into the transervse ones accompanying the
sunspot rotation. During five days' rotation of the sunspot, the
X3.4 flare erupted as mentioned above.

As is well-known, the rapid evolution of magnetic loops generally
induces evolution of the coronal magnetic fields correspondingly. A
sequence of XRT images in Fig. 5 show the evolution of coronal loops
of high temperature during the sunspot rotation. One can see the
formation process of the sheared loops structure from December 10 to
December 11. The three dashed lines in Fig. 5b denoted the three
shearing loops. An inverse S-shaped structure appeared on December
12 accompanying the sunspot rotation (see Fig. 5c). The inverse
S-shaped loop and another loop are signed by 1 and 2 respectively in
Fig. 5c. Before the X3.4 flare, the two loops almost merged into one
sheared loop denoted by the dashed curves in Fig. 5d. During the
X3.4 flare, the loops became more sheared at 02:20:18UT and the
width of loop became narrow pointed by a white arrow in Fig. 5e.
After the flare, the sheared loops disappeared and became potential
post-flare loops denoted by the dashed lines in Fig. 5f. The sheared
and the inverse S-shaped structures are in favor of the occurrence
of magnetic reconnection in common sense (Ji et al. 2007, Pevtsov
2002). The foot points of these loops can be seen rooted in the
magnetic condensed region. Therefore the rotation can shear these
loops correspondingly, then triggered the eruption.

From the above descriptions, one can not help thinking of this: the
sunspot rotation caused the loop footpoints moving and the magnetic
field lines shearing. The lower magnetic field lines of loops were
twisted in the photosphere and then the twist was transported into
the corona. It is worth pointing out that Gibson et al. (2002, 2004)
observed magnetic flux ropes in the corona and simulated the
rotation of the sunspot. They found that the sunspot rotation can
form an S-shaped structure in the corona. In this paper, we
identified the result again from observation.

\section{Summary and discussion}

In the above sections, we have investigated the active region NOAA
10930 from 2006 December 10 to 13 in detail. From the evolution of
this active region, we have found the rapid rotation of the small
sunspot. Furthermore, the rapid rotations took place before the X3.4
flare. By viewing the vector magnetograms of this active region, we
found the magnetic force lines highly sheared along the neutral line
following the sunspot rotation. From the X-ray images above this
region, the sheared loops and the inverse S-shaped loop in the
corona were identified. Comparing with the work of Zhang et al.
(2007), we analyzed this active region by using high spatial
resolution continuum intensity images, vector magnetograms, and XRT
images. Our purpose is to reveal the possible causality from the
rotating sunspot to the flare.

More important is that we combine these findings and trace the
causality chain from the sunspot rotation to the flare eruption by
using those high spatial resolution images. Therefore, we can
conclude that the rotation produces and transports the twist of the
loops through their legs to tops where the twist is evidenced by the
XRT observations. Thus one can get the result that the sunspot
rotation serves as the driver for both twisted magnetic loops
formation and their non-potential eruption. Thus, the chain of
causality for the X3.4 flare eruptions can be traced as follow: the
rapid rotation of the sunspot, the evolution of their transverse
magnetic fields, and the corresponding evolution of configuration of
magnetic loops in corona, and then the eruption. All of these
provide a reliable evidence that photospheric motion makes magnetic
loops twist from the photosphere through the chromosphere into the
corona, and then flares are caused.

As is also well known, the Coriolis force can make sunspots rotate
in a clockwise (counterclockwise) direction in the northern
(southern) while the differential rotation gives rise to the reverse
motion. Bao et al. (2002) analyzed several origins of twist of
magnetic flux tubes. Besides the Coriolis force and the differential
rotation, they have also analyzed the $\alpha$-effect, surface flow,
and magnetic reconnection. However, many different viewpoints exist
in the literature. For example, Brown et al. (2003) thought that the
differential rotation does not play a major role in producing
sunspot rotation. They suggested that the photospheric flow and
flux-tube emergence may be responsible for sunspots rotation. Su et
al. (2008) proposed that Lorentz force may be a possible driving
force for sunspot rotation. Following the development of the
technology and theory, the mechanisms of sunspot rotation are
expected to be solved in the future.

\section*{Acknowledgments}
The authors thank the Hinode and GOES consortia for their data.
Hinode is a Japanese mission developed and launched by ISAS/JAXA,
with NAOJ as domestic partner and NASA and STFC (UK) as
international partners. It is operated by these agencies in
co-operation with ESA and NSC (Norway). This work is supported by
the National Science Foundation of China (NSFC) under the grant
number 10673031 and 40636031, National Basic Research Program of
China 973 under the grant number G2006CB806301.

\label{lastpage}

\end{document}